\documentclass[twocolumn]{aastex631}
\usepackage[utf8]{inputenc}
\usepackage{natbib}
\setcitestyle{round}
\usepackage{amsmath,amssymb}
\usepackage{xcolor}
\usepackage[caption=false]{subfig}
\usepackage{soul}
\usepackage{enumitem}
\usepackage{xcolor}

\newcommand{\hi}{{\rm H\,{\small I}}}

\newcommand{\kms}{\ensuremath{{\rm km\,s^{-1}}}}

\newcommand{\ts}{\ensuremath{T_S}}

\newcommand{\persc}{\ensuremath{{\rm cm^{-2}}}}
\newcommand{\bighicat}{{\sc bighicat}}

\begin{document}

\title{Revisiting the Vertical Distribution of \hi{} Absorbing Clouds in the Solar Neighborhood. II. Constraints from a Large Catalog of 21~cm Absorption Observations at High Galactic Latitudes}

\correspondingauthor{Daniel R. Rybarczyk}
\email{rybarczyk@astro.wisc.edu}

\author[0000-0003-3351-6831]{Daniel R. Rybarczyk}
\affiliation{NSF Astronomy \& Astrophysics Postdoctoral Fellow, Department of Astronomy, University of Wisconsin--Madison, 475 N Charter St, Madison, WI 53703, USA}

\author[0000-0003-0640-7787]{Trey V. Wenger}
\affiliation{NSF Astronomy \& Astrophysics Postdoctoral Fellow, Department of Astronomy, University of Wisconsin--Madison, 475 N Charter St, Madison, WI 53703, USA}

\author[0000-0002-3418-7817]{Sne\v zana Stanimirovi\'c}
\affiliation{Department of Astronomy, University of Wisconsin--Madison, 475 N Charter St, Madison, WI 53703, USA}

\begin{abstract}
The cold neutral medium (CNM) is where neutral atomic hydrogen (\hi{}) is converted into molecular clouds, so the structure and kinematics of the CNM are key drivers of galaxy evolution. Here we provide new constraints on the vertical distribution of the CNM using the recently-developed \texttt{kinematic\_scaleheight} software package and a large catalog of sensitive \hi{} absorption observations. 
We estimate the thickness of the CNM in the solar neighborhood to be $\sigma_z\sim50$--$90~\mathrm{pc}$, assuming a Gaussian vertical distribution. 
This is a factor of $\sim2$ smaller than typically assumed, indicating the thickness of the CNM in the solar neighborhood is similar to that found in the inner Galaxy, consistent with recent simulation results.
If we consider only structures with \hi{} optical depths $\tau>0.1$ or column densities $N(\hi{})>10^{19.5}~\persc{}$, which recent work suggests are thresholds for molecule formation, we find $\sigma_z\sim50~\mathrm{pc}$. Meanwhile, for structures with $\tau<0.1$ or column densities $N(\hi{})<10^{19.5}~\persc{}$, we find $\sigma_z\sim120~\mathrm{pc}$. These thicknesses are similar to those derived for the thin- and thick-disk molecular cloud populations traced by CO emission, possibly suggesting that cold \hi{} and CO are well-mixed.
Approximately $20\%$ of CNM structures are identified as outliers, with kinematics that are not well-explained by Galactic rotation. We show that some of these CNM structures --- perhaps representing intermediate velocity clouds --- are associated with the Local Bubble wall. 
We compare our results to recent observations and simulations, and we discuss their implications for the multiphase structure of the Milky Way's interstellar medium.

\end{abstract}

\section{Introduction} \label{sec:intro}

Neutral atomic hydrogen (\hi{}) is a key ingredient in galaxy evolution. \hi{} plays a crucial role in the transition from hot ionized gas to cold molecular gas, thereby mediating star formation and stellar feedback processes (see reviews by \citealt{DickeyLockman1990}, \citealt{KalberlaKerp2009}, and \citealt{McG2023}, and references therein).
The distribution and kinematics of \hi{} set the stage for the formation of molecular clouds and stars. They are also, in turn, affected by feedback.

The balance of heating and cooling in galaxies implies that \hi{} should exist in a multiphase medium, with a colder, denser phase (the cold neutral medium, CNM) and a warmer, more diffuse phase (the warm neutral medium, WNM) that are thermally stable and exist in a rough pressure equilibrium \citep[e.g.,][]{McKeeOstriker1977}. In the Milky Way, the CNM and WNM have been characterized through observations of the 21~cm transition of \hi{} in emission and absorption \citep[][and references therein]{DickeyLockman1990,KalberlaKerp2009,McG2023}.
The distribution of atomic gas between the CNM and WNM is important because it impacts galaxy evolution. For example, recent work has revealed that only the coldest, optically thickest \hi{}, all of which is in the CNM, is associated with the formation of most molecular gas \citep{Stanimirovic2014,Nguyen2019,Ryb22,Park2023,Hafner2023}, consistent with theoretical expectations \citep[e.g.,][]{Goldsmith2007}. 
Characterizing the structure and kinematics of the CNM in the Milky Way is therefore essential to understanding the evolution of the Galactic interstellar medium (ISM).

Moreover, the structure of the multiphase ISM is shaped by myriad physical processes.
In particular, the vertical structure of the ISM is set by a balance between the gravitational force and the pressure gradient, which itself is affected by thermal, turbulent, radiative, magnetic, cosmic ray, and feedback processes in the ISM \citep[e.g.,][]{Parker1969,Bloemen1987,BoularesCox1990,LockmanGehman1991,Hill2012}. Observational constraints on the vertical distribution of the multiphase ISM are necessary to test models and simulations of galaxy evolution.

Yet, characterizing the structure of multiphase \hi{} in the Galaxy has remained difficult. For example, observations of \hi{} absorption are needed to unambiguously detect the CNM, but such observations are limited to a relatively small number of sightlines in the direction of randomly-distributed background radio continuum sources. 
\citet{Crovisier1978} developed a novel statistical technique for inferring the vertical thickness of the CNM disk using the positions and radial velocities of a sample of \hi{} structures identified in absorption.
The observed velocity of an \hi{} structure is a combination of Galactic rotation, the sun's motion relative to the local standard of rest (LSR), and a random component. \citet{Crovisier1978} expressed the Galactic rotation component in terms of the mean displacement of \hi{} structures from the plane in the vertical ($z$) direction, $\langle |z| \rangle$.
Then, by minimizing the difference between the expected and observed radial velocities, they were able to constrain $\langle |z| \rangle$ (see Section \ref{sec:methods}).
They applied this technique to \hi{} absorption observations (ensuring they probed just the CNM) at high Galactic latitudes ($|b|\gtrsim10^{\circ}$, where discrete spectral features in the absorption spectra can reliably be identified; e.g., \citealt{Murray2017}), tracing primarily local gas structures ($d\lesssim1~$kpc). 
For their sample of $\sim300$ absorbing \hi{} structures \citep{Crovier1978_survey}, they estimated $\langle |z| \rangle=(107\pm29)~\mathrm{pc}$. For a Gaussian distribution, this corresponds to standard deviation in the $z$ direction of $\sigma_z=(134\pm~36)~\mathrm{pc}$.

But, this result was recently called into question when \citet{Wenger2024} identified an error made by \citet{Crovisier1978}. \citet{Wenger2024} showed that, for any sample of structures truncated in latitude (including the sample used by \citealt{Crovisier1978}, with a threshold $|b|>10^\circ$), the quantity that \citet{Crovisier1978} measured was not, in fact, the mean displacement of the vertical distribution, but was instead a ratio of higher moments of the vertical distribution. Since \citet{Crovisier1978} \citep[and later][]{BelfortCrovisier1984} reported this quantity as the mean displacement, the vertical thickness of the CNM in the solar neighborhood was overestimated. \citet{Wenger2024} developed an updated approach that accounts for distributions that are truncated in latitude. Using the same sample as \citet{Crovisier1978}, they found $\sigma_z=61.7^{+9.6}_{-9.0}~\mathrm{pc}$ (assuming a Gaussian distribution), a factor of $\sim2$ lower than the reported \citet{Crovisier1978} result.

Meanwhile, other approaches have been used to measure the thickness of the CNM at different Galactocentric radii. \citet{Dickey2009} used observations of \hi{} absorption from three Galactic plane surveys \citep{Taylor2003,McG2005,Stil2006} to estimate the thickness of the CNM in the outer Galaxy, and \citet{Dickey2022} used observations of \hi{} absorption from the Australian Square Kilometre Array Pathfinder \citep[ASKAP;][]{Hotan2021} to estimate the thickness of the CNM in the inner Galaxy. These projects characterized absorption at discrete Galactocentric radii (derived from the kinematics) rather than from discrete absorbing components, as the spectra in the plane are too complex for reliable Gaussian decomposition. In the inner Galaxy ($R\sim2.8$--$3.7~\mathrm{kpc}$), \citet{Dickey2022} found a CNM thickness of $\sigma_z\sim50$--$90~\mathrm{pc}$. At the solar circle (taken on the far side of the Galaxy, at a distance of $15.6~\mathrm{kpc}$), they estimated $\sigma_z=160~\mathrm{pc}$. In the outer Galaxy (from $R\sim8.5$--$25~\mathrm{pc}$), \citet{Dickey2009} found that the CNM had a thickness $\sigma_z\sim170$--$300~\mathrm{pc}$ (see their Figure 7), showing the flaring of the disk beyond the solar circle. They found that the thickness of the CNM at these radii was comparable to the thickness of the WNM. 

\citet{McG2023} summarized the previous estimates of the CNM thickness as a function of Galactocentric radius (their Figure 11), showing a gradual rise in the CNM thickness from the inner to the outer Galaxy, similar to the trends seen for the WNM \citep{Levine2006,KalberlaKerp2009} and for the molecular gas \citep[][and references therein]{HeyerDame2015}. They noted that this picture is consistent with expectations given the higher pressure in the inner Galaxy. However, when the CNM thickness in the solar neighborhood value is updated from the \citet{Crovisier1978} result to the new result from \citet{Wenger2024}, a different picture emerges. In contrast to the paradigm where the CNM thickness increases with increasing Galactocentric radius, this new result suggests that the thickness of the CNM in the solar neighborhood is actually comparable to that found in the inner Galaxy, at a Galactocentric radius of $\sim3~\mathrm{kpc}$ \citep{Dickey2022}.

Here, we extend the work done by \citet{Wenger2024} by applying the \texttt{kinematic\_scalehieght} code they developed \citep{Wenger2024_code} to a large catalog of high-latitude \hi{} absorption observations compiled by \citet{McG2023}, the \bighicat{} (Section \ref{sec:data}). Besides refining the method developed by \citet{Crovisier1978} for observations truncated in latitude, \citet{Wenger2024} further developed two new Bayesian models that use Monte Carlo Markov Chain (MCMC) methods to constrain the thickness of the disk in a more reliable way (Section \ref{sec:methods}). In Section \ref{sec:results}, we derive the vertical thickness of the CNM disk using each of these methods to different samples of the \bighicat{}. We present not only new estimates of the vertical thickness of the CNM, but also new estimates of the cloud-to-cloud velocity dispersion and constraints on the solar motion with respect to the LSR. We also briefly investigate the properties of absorbing features with anomalous velocities that are identified as outliers. In Section \ref{sec:discussion}, we discuss our results in the context of previous estimates of the CNM disk thickness as well as recent galaxy-scale simulations of the multiphase ISM. Finally, we present our conclusions in Section \ref{sec:conclusions}.

\section{Data}\label{sec:data}

\begin{figure*}[!ht]
    \centering
    \includegraphics[width=\linewidth]{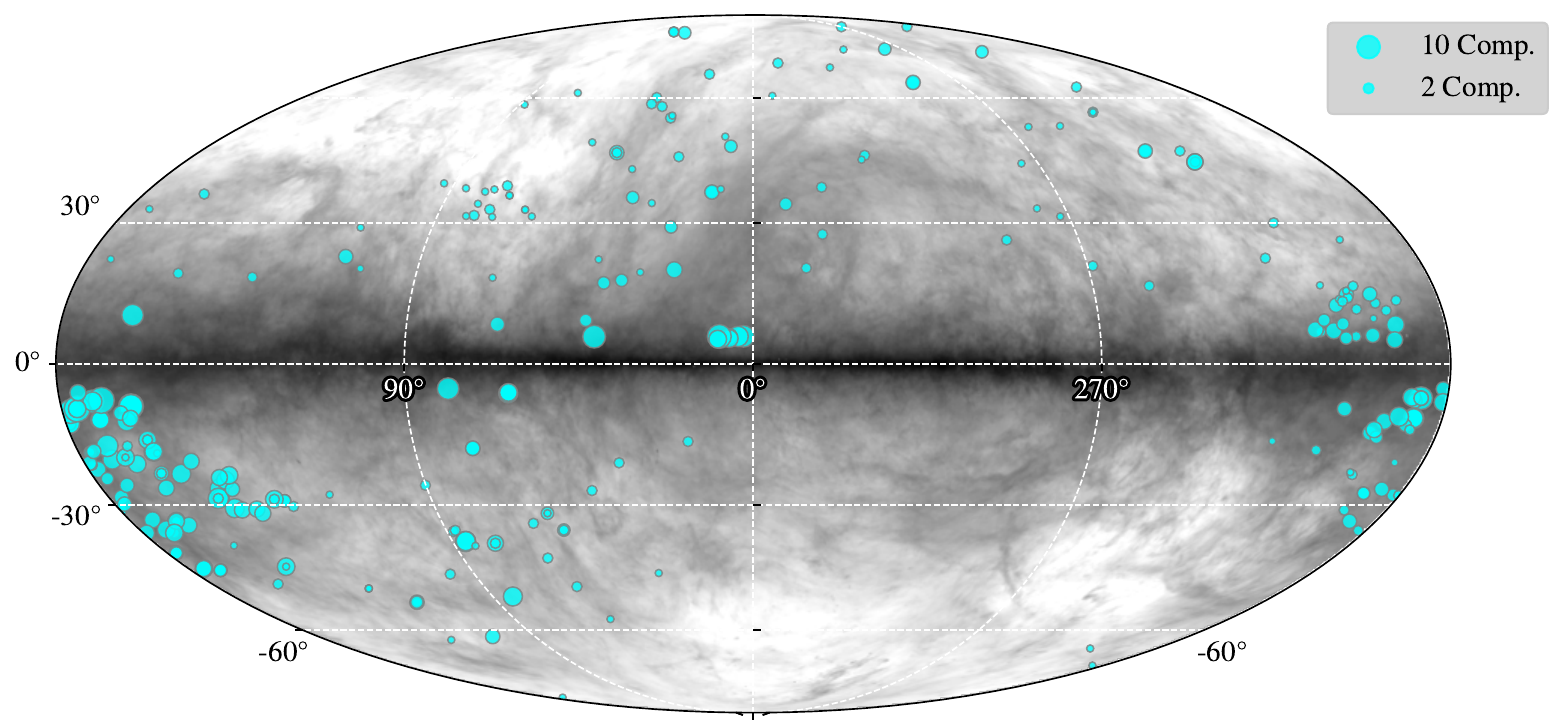}
    \includegraphics[width=\linewidth]{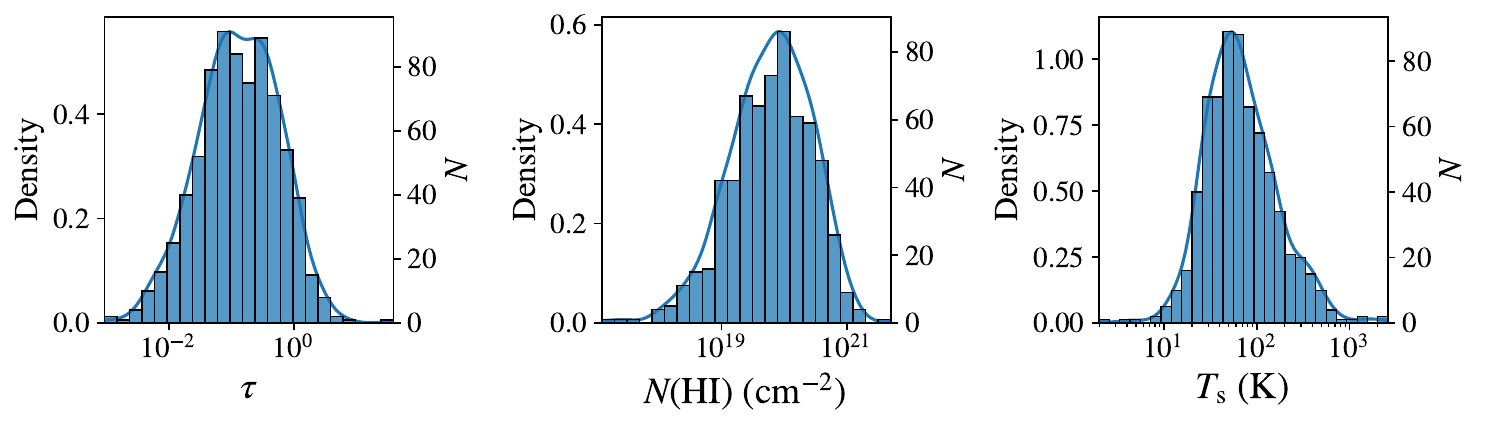}
    \caption{\textit{Top:} The positions of of \bighicat{} background sources at $|b|>5^\circ$, sized according to the number of components detected along each line of sight. The background is a Mollweide projection of \hi{} emission integrated from $-75~\kms{}$ to $+75~\kms{}$ \citep{HI4PI}. \textit{Bottom:} distribution of \hi{} optical depth (left), the \hi{} column density (center), and the \hi{} spin temperature (right) for structures in the \bighicat{} sample with $|b|>5^\circ$. In each panel, a histogram is shown with the kernel density estimation overlaid.}
    \label{fig:BIGHICAT_HI_properties}
\end{figure*}
\citet{McG2023} compiled \hi{} absorption data from seven 21~cm line surveys --- the Millennium Arecibo 21~cm Absorption-Line Survey with the Arecibo Observatory \citep{Heiles2003a,Heiles2003b}, a high latitude survey with the Giant Metrewave Radio Telescope \citep{Mohan2004a,Mohan2004b}, a survey in the direction of the Perseus molecular cloud with Arecibo \citep{Stanimirovic2014}, the 21~cm Spectral Line Observations of Neutral Gas with the Karl G. Jansky Very Large Array (VLA) \citep[21-SPONGE;][]{Murray2015,Murray2018}, a survey of \hi{} absorption in the direction of five giant molecular clouds (Taurus, California, Rosette, Mon OB1, and NGC 2264) with Arecibo \citep{Nguyen2019}, and the Measuring Absorption by Cold Hydrogen (MACH) survey with the VLA \citep{Murray2021} --- into a catalog called the \bighicat{}. 
The \bighicat{} comprises 1223 unique Gaussian components identified in \hi{} absorption. 

Here we consider the 768 \bighicat{} features identified at Galactic latitudes $|b|>5^\circ$. Following \citet{Crovisier1978}, we focus only on higher latitudes because the absorption spectra are simpler and can be more reliably decomposed into Gaussian components \citep[e.g.,][]{Murray2017}. Moreover, the thickness of the \hi{} disk is thought to be a function of Galactocentric radius \citep[e.g.,][]{Levine2006,KalberlaKerp2009}. Focusing on high latitude observations allows us to trace primarily local structures ($d<2~\mathrm{kpc}$), whereas features detected closer to the plane may probe different regions of the Galaxy.
In Figure \ref{fig:BIGHICAT_HI_properties}, we show the distribution of the optical depths ($\tau$), \hi{} column densities ($N(\hi{})$), and spin temperatures\footnote{The spin temperature of H\,{\scriptsize I} is the excitation temperature for the 21~cm hyperfine transition.} ($\ts$) for the total sample. While the optical depth is measured for all components, only $87\%$ of components have estimated spin temperatures and $83\%$ of components have estimated column densities (see \citealt{McG2023} for discussion of the \bighicat{} construction and completeness).

\section{Methods} \label{sec:methods}
We apply the \texttt{kinematic\_scaleheight} code \citep{Wenger2024_code} developed by \citet{Wenger2024} to different subsamples of the \bighicat{} to characterize the vertical distribution of the CNM in the solar neighborhood.
As in \citet{Crovisier1978}, this algorithm infers the mean vertical displacement by comparing the observed radial velocities of CNM clouds with the expected radial velocities,
\begin{equation}
    V_{\mathrm{LSR}} = V_{\odot,\mathrm{LSR}}(l,b) + V_{\mathrm{rot}}(d,l,b) + V_t,
\end{equation}
where $V_{\odot,\mathrm{LSR}}$ is the line-of-sight velocity of the sun with respect to the LSR, $V_{\mathrm{rot}}$ is the line-of-sight velocity of the cloud due to Galactic rotation, and $V_t$ is the line-of-sight velocity of the cloud due to random cloud-to-cloud motions.
\citet{Crovisier1978} expressed $V_{\mathrm{rot}}$ in terms of $\langle |z|\rangle$ and used a least-squares method to infer $\langle |z|\rangle$. However, \citet{Wenger2024} identified an error in their method, showing that the quantity that \citet{Crovisier1978} measured was not $\langle |z| \rangle$, but instead a ratio of the higher moments of the $z$ distribution.
The \texttt{kinematic\_scaleheight} code implements a least-squares method that infers this moment ratio (which, contrary the assumption of \citealt{Crovisier1978}, is not equal to $\langle |z| \rangle$).
The \texttt{kinematic\_scaleheight} code also implements two Bayesian models --- the ``moment ratio'' method parameterized in terms of $\langle|z|^3\rangle/\langle|z|^2\rangle$ and the ``shape'' method parameterized in terms of a shape parameter (which assumes a shape of the vertical distribution) --- that implement MCMC methods to infer the posterior distributions of model parameters. Both of the MCMC methods also identify outliers and infer model parameters only for the non-outlying data.

\begin{figure*}
    \centering
    \includegraphics[width=0.9\linewidth]{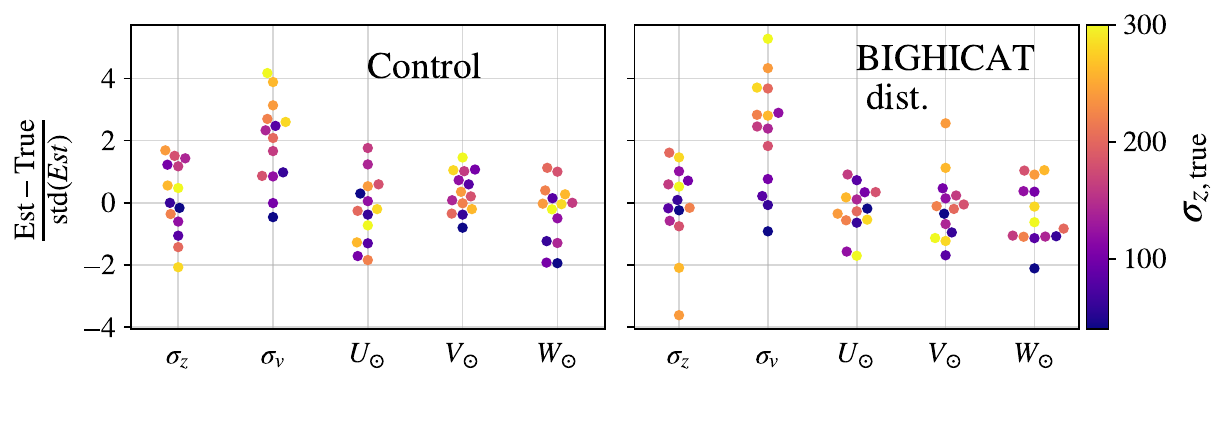}
    \caption{Swarm plots of the parameters derived for synthetic populations of clouds drawn randomly across the sky (left, ``Control'') and synthetic populations of clouds drawn from the same $(\ell,b)$ distribution as the \bighicat{} background sources (right, ``\bighicat{} dist.''). The $x$-axis indicates the variable ($\sigma_z$, $\sigma_v$, $U_{\odot}$, $V_{\odot}$, or $W_{\odot}$; small horizontal offsets are used to prevent points from overlapping) and the $y$-axis indicates the error in the estimated parameter, normalized by the uncertainty in the fit. We show results for 14 synthetic samples, each with different values of $\sigma_z$, ranging from $40~\mathrm{pc}$ to $300~\mathrm{pc}$; points are colored according to the value of $\sigma_z$ for each synthetic sample.}
    \label{fig:bighicat_bias}
\end{figure*}

When applying the \texttt{kinematic\_scaleheight} code to the \bighicat{}, we must consider whether the distribution of background sources could be biasing our results.
The \bighicat{} background sources are not randomly distributed across the sky --- the fraction of components at $|b|>5^\circ$ in the first, second, third, and fourth Galactic quadrants are $26.6\%$, $43.1\%$, $3.6\%$, and $26.7\%$, respectively. 
Meanwhile, $49.0\%$ of those components have latitudes $5^\circ\leq|b|\leq20^\circ$, $25.7\%$ have latitudes $20^\circ<|b|\leq35^\circ$, $17.1\%$ have latitudes $35^\circ<|b|\leq50^\circ$, and $8.3\%$ have latitudes $|b|>50^\circ$.
To test how this non-uniform sky coverage affects the fitting results, we apply the fitting to synthetic populations of clouds with different values of $\sigma_z$.
For each synthetic population of clouds, we apply the \texttt{kinematic\_scaleheight} code to a subsample of 200 clouds (comparable to the smallest sample size in Table \ref{tab:fitting_results}) drawn from the same longitude-latitude distribution as the \bighicat{} sample, and to another subsample of 200 randomly-selected clouds. 
For samples with $\sigma_z\leq150~\mathrm{pc}$, we find that the two Bayesian techniques converge on the correct solutions for $\sigma_z$, $\sigma_v$ (see Section \ref{subsec:sigma_v}), $U_{\odot}$, $V_{\odot}$, and $W_{\odot}$ (within $\lesssim2.5\sigma$) in all cases, regardless of the sky distribution of the clouds.\footnote{The results for the least-squares technique similarly converge to the correct solution (within $2.5\sigma$) for $U_{\odot}$, $V_{\odot}$, and $W_{\odot}$, but sometimes overstimate $\sigma_z$ and $\sigma_v$ by a larger degree than the Bayesian methods.} In Figure \ref{fig:bighicat_bias}, we show swarm plots for the error in the derived parameters for the control (randomly-distributed) samples and the samples drawn from the same distribution as the \bighicat{}. We specifically present results from the shape model, but the results for the two MCMC techniques are not statistically significantly different.  We show results for 14 synthetic samples, each with different values of $\sigma_z$, ranging from $40~\mathrm{pc}$ to $300~\mathrm{pc}$; points are colored according to the value of $\sigma_z$ for each synthetic sample. These swarmplots show that (1) the non-random distribution of the clouds in the sample does not strongly bias our results and (2) for both the randomly and non-randomly drawn samples, the Bayesian methods converge on the correct solution. The only exception is that we tend to slightly underestimate $\sigma_v$, and this bias is slightly enhanced in the \bighicat{}-distribution samples.  For samples with $\sigma_z\lesssim150~\mathrm{pc}$ \citep[which we expect for the local CNM, e.g.,][]{Dickey2022,Wenger2024}, this bias is $<3\sigma$, but for some samples with larger $\sigma_z$, the bias is as high as $\sim4\sigma$.
We take the results of these tests as evidence that the distribution of \bighicat{} background sources should not significantly impact our estimates of the CNM thickness in the solar neighborhood.

Finally, we argue that our results are unlikely to be affected by the presence of a Galactic lagging halo. While a $z$-dependent velocity lag could significantly bias the results (since both \citealt{Crovisier1978} and \citealt{Wenger2024} ignore such a lag), typical velocity lags of \hi{} in disc galaxies are  $\sim-10~\kms{}~\mathrm{kpc}^{-1}$ \citep[e.g.,][]{Marasco2019}, and for the Milky Way in particular, \citet{MarascoFraternali2011} found the lag to be $-15\pm4~\kms{}~\mathrm{kpc}^{-1}$. In Section \ref{sec:results}, we find that most of the \hi{} exists within approximately $\pm100~\mathrm{pc}$ of the plane, so we expect that any velocity offsets introduced by a lagging halo are significantly smaller than the cloud-to-cloud velocity dispersion of cold \hi{} \citep[see Section \ref{subsec:sigma_v} and, e.g.,][]{Crovisier1978,BelfortCrovisier1984,McGDickey2007,Wenger2024}.

\section{Results} \label{sec:results}
\subsection{The thickness of the CNM in the solar neighborhood}
In Table \ref{tab:fitting_results}, we report the derived vertical thicknesses for different subsamples of the \bighicat{} sample.
The results in Table \ref{tab:fitting_results} differ significantly between the different samples --- the vertical distribution of \hi{} is not the same for all samples of absorbing structures. Here we investigate how the thickness of the cold \hi{} disk depends on the atomic gas properties. 
For consistency, we assume a Gaussian distribution\footnote{As in \citet{Wenger2024}, we find no preference for a Gaussian, exponential, or rectangular model using leave-one-out cross-validation \citep{Vehtari2021} --- we cannot distinguish between any of these possible shapes to the vertical distribution from the \bighicat{} data.} and report the standard deviation in the $z$ direction, $\sigma_z$. We report the fits from the updated least-squares method,  the moment ratio method, and the shape method introduced by \citet{Wenger2024} ($\sigma_{z,\mathrm{ls}}$, $\sigma_{z,\mathrm{MR}}$, and $\sigma_{z,\mathrm{shape}}$, respectively).
In all cases, $\sigma_{z,\mathrm{MR}}$ and $\sigma_{z,\mathrm{shape}}$ are consistent with each other, varying by $<0.2\sigma$. 
The estimates from the corrected least squares method are higher than those from the moment ratio method and shape method by $\gtrsim3\sigma$ in two out of nine cases, emphasizing the importance of the treatment of outliers in the two MCMC methods.

\begin{deluxetable*}{|c|c|c|c|c|} \label{tab:fitting_results}
\tablecaption{Estimates of $\sigma_z$ from the \texttt{kinematic\_scaleheight} code for different samples of \bighicat{} components. The first column describes each subsample. The second column lists the number of unique \bighicat{} components, $N$, belonging to each subsample. The third column lists the vertical thickness derived using the \texttt{kinematic\_scaleheight} least-squares method, $\sigma_{z,\mathrm{ls}}$. The fourth column lists the vertical thickness derived using the \texttt{kinematic\_scaleheight} moment ratio method, $\sigma_{z,\mathrm{MR}}$. The fifth column lists the vertical thickness derived using the \texttt{kinematic\_scaleheight} shape method, $\sigma_{z,\mathrm{shape}}$. In all cases, we assume a Gaussian vertical distribution.}
\tablehead{
\colhead{Sample} & \colhead{$N$} &  \colhead{$\sigma_{z,\mathrm{ls}}$} & \colhead{$\sigma_{z,\mathrm{MR}}$} & \colhead{$\sigma_{z,\mathrm{shape}}$} \\
\colhead{} & \colhead{} &  \colhead{pc} & \colhead{pc} & \colhead{pc}
}
\startdata
    $|b|>5^{\circ}$ & 768 & $98.9 \pm 7.6$ & $74.3 \pm 7.8$ & $75.0 \pm 8.0$ \\
    $\tau>0.1, |b|>5^{\circ}$ & 431 & $81.0 \pm 7.1$ & $58.7 \pm 7.0$ & $59.2 \pm 7.0$ \\
    $\tau<0.1, |b|>5^{\circ}$ & 336  & $123.5 \pm 14.7$ & $119.8 \pm 15.5$ & $122.4 \pm 15.3$ \\
    $N(\mathrm{HI})>10^{19.5}~\mathrm{cm}^{-2}, |b|>5^{\circ}$ & 429  & $71.0 \pm 7.5$ & $47.7 \pm 6.3$ & $48.0 \pm 6.4$ \\
    $N(\mathrm{HI})<10^{19.5}~\mathrm{cm}^{-2}, |b|>5^{\circ}$ & 216 & $152.7 \pm 19.9$ & $127.2 \pm 22.9$ & $133.7 \pm 23.6$ \\
    $T_{\mathrm{s}}<80~\mathrm{K}, |b|>5^{\circ}$ & 423 & $110.9 \pm 11.8$ & $61.4 \pm 9.8$ & $62.8 \pm 9.8$ \\
    $T_{\mathrm{s}}>80~\mathrm{K}, |b|>5^{\circ}$ & 257 & $74.9 \pm 11.4$ & $57.5 \pm 11.2$ & $59.0 \pm 11.4$ \\
    $\ts{}<80~\mathrm{K},\tau>0.1, |b|>5^{\circ}$ & 296  & $80.4 \pm 10.0$ & $46.0 \pm 7.2$ & $46.5 \pm 7.3$ \\
    $\ts{}>80~\mathrm{K},\tau<0.1, |b|>5^{\circ}$ & 171 & $115.9 \pm 18.0$ & $107.5 \pm 21.3$ & $111.9 \pm 20.9$ \\
\enddata
\end{deluxetable*}

\subsubsection{Optical depth dependence} \label{subsec:optical_depth}
The \hi{} optical depth is proportional to the density of atomic hydrogen and inversely proportional to its spin temperature and velocity dispersion. Previous work has shown that most of the molecule formation in the diffuse ISM is associated with only the optically thickest \hi, $\tau\gtrsim0.1$  \citep[][]{Stanimirovic2014,Nguyen2019,Ryb22,Park2023,Hafner2023}.

In Table \ref{tab:fitting_results}, we report that $\sigma_z=59.2\pm7.0$ for the sample of \bighicat{} structures with $\tau\geq0.1$, while $\sigma_z=122.4\pm15.3$ for the sample of structures with $\tau<0.1$.
The estimate for the optically thicker sample is in good agreement with estimates for the thickness of the molecular gas disk at distances $\lesssim2~\mathrm{kpc}$ from the sun, $\sigma_z\approx40$--$70~\mathrm{pc}$ \citep{Sanders1984,Grabelsky1987,Bronfman1988,Clemens1988,Malhotra1994,Nakanishi2006}.
Meanwhile, the estimate for the optically thinner sample is closer to what is estimated for \hi{} seen in emission. In the solar neighborhood, the thickness of the disk derived from \hi{} emission observations is $\sigma_z\approx110$--$180~\mathrm{pc}$ \citep{FalgaroneLequeux1973,KalberlaDedes2008,KalberlaKerp2009}. Many authors have used a two-component fit to account for the vertical distribution of the \hi{} seen in emission, with $\sigma_{z,1}\approx100~\mathrm{pc}$ and $\sigma_{z,2}\approx240~\mathrm{pc}$ \citep{Lockman1986,DickeyLockman1990,LockmanGehman1991}. Our result for the $\tau<0.1$ sample is similar to the thickness of the thinner of the two components in such models.

\begin{figure}
    \centering
    \includegraphics[width=\columnwidth]{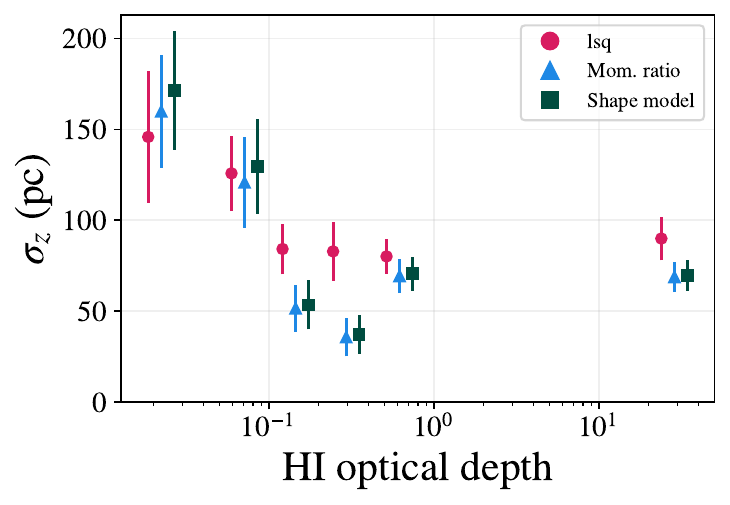}
    \caption{Fits to $\sigma_z$ for six subsamples of the \bighicat{} sorted by \hi{} optical depth. Results for the updated least-squres, the moment ratio, and the shape methods implemented by the \texttt{kinematic\_scaleheight} code are shown. The $x$-axis value for each point is placed near the center of the optical depth range of the corresponding subsample; the $x$-axis value for each method slightly offset for clarity.}
    \label{fig:sigz_v_tau}
\end{figure}

In Figure \ref{fig:sigz_v_tau}, we take a more fine-grained approach by considering six samples, each with $\sim125$ features, sorted by increasing optical depth. 
The thicknesses we measure for two samples comprising \hi{} structures with $\tau<0.1$ are both $>100~\mathrm{pc}$, while the thicknesses we measure for the four samples comprising \hi{} structures  with $\tau>0.1$ are all around $\sigma_z\approx50\text{--}70~\mathrm{pc}$.
The optical depth threshold of 0.1 presented in Table \ref{tab:fitting_results} was motivated by the fact that this appears to be a necessary (but not sufficient) criterion for molecule formation in the diffuse ISM \citep{Nguyen2019,Ryb22,Park2023,Hafner2023}. Evidently, a similar threshold also delineates cold \hi{} clouds in terms of their vertical distributions: the optically thicker sample is more tightly confined to the plane (intriguingly, in a vertical distribution similar to that of the molecular gas in the solar neighborhood), while the optically thinner sample has a wider distribution in the $z$ direction (closer to that derived from \hi{} emission).

\subsubsection{\hi{} column density dependence} \label{subsec:NHI}
The column density of an atomic gas structure is related to the density of the structure, $N(\hi{})=\int n(\hi{})~dl$, and indicates how well-shielded the gas is (although this depends on the geometry of the gas structure). \citet{Ryb22} showed that diffuse molecular gas was associated only with CNM structures that had a column density $N(\hi{})\gtrsim10^{19.5}$.
In Table \ref{tab:fitting_results}, we report that $\sigma_z=133.7\pm23.6$ for structures in the \bighicat{} with $N(\hi)<10^{19.5}~\persc$, while $\sigma_z = 48.0\pm6.4$ for the structures with $N(\hi)>10^{19.5}~\persc{}$.
These results are consistent with the vertical distributions inferred for the $\tau < 0.1$ and $\tau > 0.1$ samples, respectively. The \hi{} optical depth and \hi{} column density are well correlated, and 74\% of structures with $N(\hi{})>10^{19.5}~\persc{}$ have $\tau>0.1$, so it is unsurprising that the results are in good agreement with the optical depth results discussed above.

\begin{figure}
    \centering
    \includegraphics[width=\columnwidth]{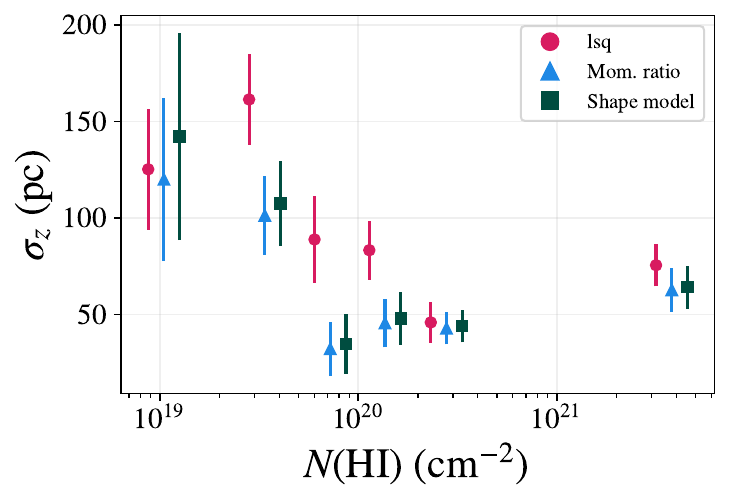}
    \caption{Same as Figure \ref{fig:sigz_v_tau}, but for six subsamples of the \bighicat{} sorted by \hi{} column density.}
    \label{fig:sigz_v_NHI}
\end{figure}

We take a more fine-grained look at the dependence of $\sigma_z$ on \hi{} column density in Figure \ref{fig:sigz_v_NHI}. We sort the \bighicat{} by \hi{} column density and consider six samples, each with $\sim105$ structures. Figure \ref{fig:sigz_v_NHI} shows that the  thicknesses we measure for two samples comprising \hi{} structures with $N(\hi{})\lesssim10^{20}~\persc{}$ are both $>100~\mathrm{pc}$, while the thicknesses we measure for the four samples comprising \hi{} structures  with $N(\hi{})\lesssim10^{20}~\persc{}$ are all around $\sigma_z\approx50\text{--}70~\mathrm{pc}$.
These results are very similar to those we found for the \hi{} optical depth above, but this is again unsurprising since correlation between the \hi{} optical depth and \hi{} column density suggests that the two analyses are probing similar structures.
Here, we find that a threshold column density of $\sim10^{20}~\persc{}$ delineates the thinner and thicker \hi{} populations discussed above.

\subsubsection{Temperature dependence} \label{subsec:temperature}
The spin temperature of \hi{} is approximately equal to its kinetic temperature in the high density CNM, where the 21~cm transition is thermalized by collisions with electrons, ions, and other hydrogen atoms.
In the high latitude \bighicat{} sample used here, 91\% of features have $\ts<250~\mathrm{K}$ and only four of the 680 measured spin temperatures have $\ts>1000~\mathrm{K}$. 
We therefore treat \ts{} in this sample as a reasonable estimate of the kinetic temperature, allowing us to investigate how the thickness of the disk depends on the temperature of the atomic gas.
In the solar neighborhood, the CNM is expected to have temperatures $\ts{}\lesssim250~\mathrm{K}$, while the WNM is expected to have temperatures $\ts\gtrsim 4000~\mathrm{K}$ \citep[e.g.,][]{Wolfire2003}. Gas in the intermediate temperature regime belongs to the unstable neutral medium (UNM), which is thermally unstable but which comprises about 20\% of the total mass of \hi{} in the Milky Way \citep{Murray2018}.
Comparisons of \hi{} absorption measurements to molecular gas observations have confirmed that only the coldest atomic gas is associated with molecule formation --- \citet{Ryb22} found molecular gas associated only with structures that had $\ts\lesssim80~\mathrm{K}$, while \citet{Park2023} found a threshold of $\ts\lesssim200~\mathrm{K}$ and \citet{Hafner2023} found $\ts\lesssim140~\mathrm{K}$).

\begin{figure}
    \centering
    \includegraphics[width=\columnwidth]{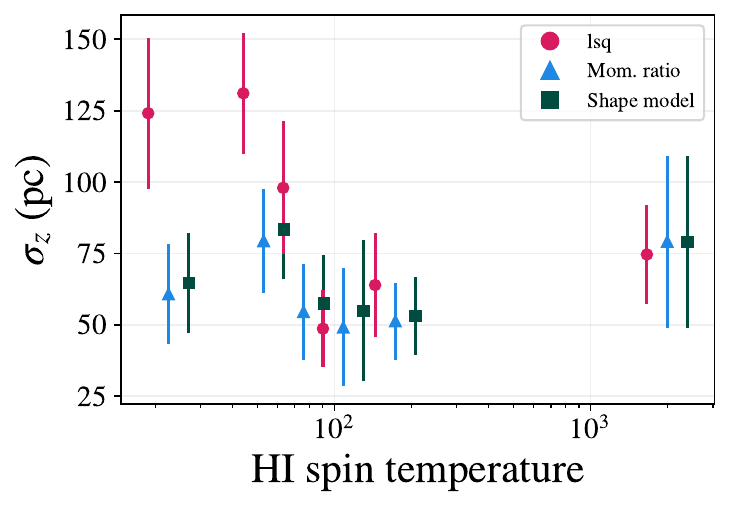}
    \caption{Same as Figure \ref{fig:sigz_v_tau}, but for six subsamples of the \bighicat{} sorted by \hi{} spin temperature.}
    \label{fig:sigz_v_Ts}
\end{figure}

In Table \ref{tab:fitting_results}, we report that $\sigma_z=59.0\pm11.4$ for the sample of \bighicat{} structures with $\ts>80~\mathrm{K}$, where it is unlikely for \hi{} to be associated with molecule formation \citep[e.g.,][]{Ryb22}. Meanwhile, for the sample of structures with $\ts<80~\mathrm{K}$, we find $\sigma_z=62.8\pm9.8$. Whereas the optical depth and column density thresholds for molecule formation \citep{Nguyen2019,Ryb22,Park2023,Hafner2023} seem to be associated with changes in the vertical structure of the \hi{}, we do not see such a delineation associated with the temperature threshold.
In Figure \ref{fig:sigz_v_Ts}, we further divide the sample into six bins each with $\sim110$ components sorted by increasing temperature. Again, we find $\sigma_z\approx60$--$80~\mathrm{pc}$ in all samples, with no statistically significant variations between any of the six samples.

\citet{Hill2018} investigated the vertical distribution of gas  as a function of temperature in 3D magnetohydrodynamic simulations of the multiphase ISM \citep{JoungMacLow2006,Joung2009,Hill2012}. Their Figure 6 suggests that the atomic gas at temperatures between $T\sim20~\mathrm{K}$ and $T\sim200~\mathrm{K}$ is fairly uniformly distributed in the range $0~\mathrm{pc} \lesssim |z|\lesssim100~\mathrm{pc}$. We note that 89\% of the high-latitude \bighicat{} sample  has $\ts<200~\mathrm{K}$.

In Table \ref{tab:fitting_results}, we also consider the optical depth and spin temperature together, as both seem to be important to setting the conditions for molecule formation \citep[e.g.,][]{Ryb22,Park2023,Hafner2023}. We find that for structures with $\ts<80~\mathrm{K}$ and $\tau>0.1$ --- conditions necessary but not sufficient for the formation of diffuse molecular gas in the sample of \citet{Ryb22} --- the measured thickness is $\sigma_z=46.5\pm7.3$, while for structures with $\ts>80~\mathrm{K}$ and $\tau<0.1$, the measured thickness is $\sigma_z=111.9\pm20.9$. These results are not significantly different from those where we considered only the optical depth threshold.

\subsubsection{Comparison to other datasets}
\begin{figure*}
    \centering
    \includegraphics[width=0.7\linewidth]{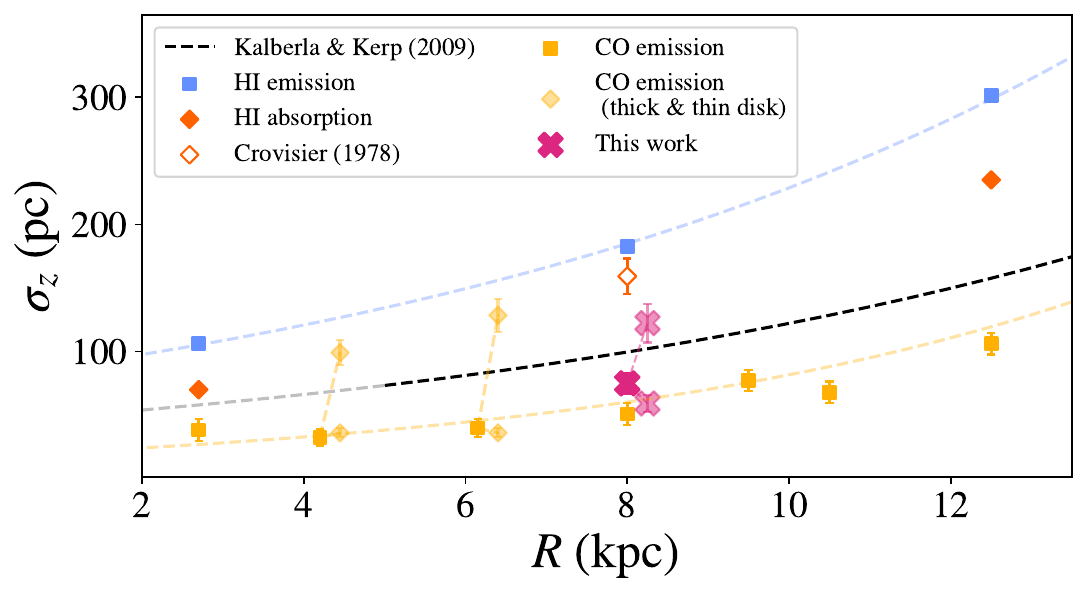}
    \caption{A recreation of Figure 11 of \citet{McG2023}, with new estimates of $\sigma_z$ derived in this work overplotted. Yellow squares indicate the thickness of the molecular gas disk derived from observations of CO in emission \citep[see Figure 6][and references therein]{HeyerDame2015}. For the Galactocentric radii probed by \citet{Su2021}, we separate $\sigma_z$ into the thick and thin components using dashed lines, with semi-transparent yellow squares indicating $\sigma_z$ for the molecular thin and thick disks. Orange diamonds indicate the thickness of the cold \hi{} disk derived from observations of \hi{} in absorption \citep{Dickey2009,Dickey2022}. The solar neighborhood estimate for the cold \hi{} disk from \citet{Crovisier1978} used by \citet{McG2023} is shown as an unfilled orange diamond. Blue squares indicate the thickness of the total \hi{} disk derived from observations of \hi{} in emission \citep{HI4PI}. Estimates from this work (in the solar neighborhood; Table \ref{tab:fitting_results}) are shown as red X's. The result for the full $|b|>5^\circ$ sample is shown as an opaque X. We then separate $\sigma_z$ for cold \hi{} in the solar neighborhood into the thin and thick components (corresponding to results for the $\tau>0.1$ and $\tau<0.1$ samples; Table \ref{tab:fitting_results}) using dashed lines, with semi-transparent X's indicating $\sigma_z$ for these two samples. A black dashed line shows the exponential fit to the \hi{} disk thickness from \citet{KalberlaKerp2009}, valid for Galactocentric radii $5~\mathrm{kpc} \lesssim R \lesssim 35~\mathrm{kpc}$ (a gray dashed line shows the extrapolated fit for $R<5~\mathrm{kpc}$). Yellow and blue dashed lines show our exponential fits to the disk thickness determined from \hi{} emission and CO emission, respectively.}
    \label{fig:araa_comparison}
\end{figure*}

In Figure \ref{fig:araa_comparison}, we recreate Figure 11 of \citet{McG2023}, which shows the thickness of the disk as a function of Galactocentric radius measured in \hi{} emission (blue squares), \hi{} absorption (orange diamonds), and CO emission (yellow squares). We overplot results from this work in pink X's. The thickness measured for the total high-latitude sample is shown as a dark X, while those measured for the sample with $\tau>0.1$ and  $\tau<0.1$ are showed as semi-transparent X's (these values come from the first three columns of Table \ref{tab:fitting_results}).
The result from \citet{Crovisier1978} is shown as an open diamond. As discussed above, the optically thicker sample has a thickness close to that measured for the molecular gas disk. The optically thinner sample is associated with a thicker disk, about halfway between the molecular gas disk and the disk measured in \hi{} emission.

The exponential fit to the thickness of the \hi{} disk is shown as a black line (interpolated as a gray line for $R<5~\mathrm{kpc}$), with a radial scale length of $\sim9.8~\mathrm{kpc}$ \citep[][]{Kalberla2007,KalberlaDedes2008}. We further fit exponential functions to the CO \citep{Clemens1988,Bronfman1988} and WNM \citep{McG2023}, shown in yellow and blue, respectively.  For the CO, we find a radial scale length of $6.6\pm0.6~\mathrm{kpc}$, while for the WNM, we find a radial scale length of $9.4\pm1.3~\mathrm{kpc}$. We do not find a good exponential fit for the CNM ($\chi^2\gg1$), which is reasonable if the thickness of the CNM is indeed constant for much of the Galaxy, as suggested by, e.g., \citet{Smith2023} (see Section \ref{sec:discussion}). Nevertheless, it is worth noting the the CNM data are sparse and that our methods are different than those used by \citet{Dickey2009,Dickey2022}, so future work will be needed to more definitively characterize the radial structure of the CNM.

More work is needed to understand the connection between the thin and thick disks measured in CO and the thin and thick disks inferred here for the optically thicker and optically thinner \hi{}, respectively. Direct comparisons of \hi{} absorption to molecular line observations have been extremely limited due to the relatively small number of sightlines where \hi{} absorption has been measured \citep[e.g.,][]{Nguyen2019,Ryb22,Park2023,Hafner2023}. Moreover, observations of the thick molecular gas have probed Galactocentric radii less than $R_{\odot}$ \citep[derived by observing gas at tangent points in the first quadrant; ][]{DameThaddeus1994,Malhotra1994,Su2021}, whereas our observations probe the solar neighborhood (see Figure \ref{fig:araa_comparison}). Nevertheless, the intriguing similarity may suggest that cold \hi{} and molecular gas are well-mixed.

\subsection{The velocity dispersion of cold \hi{}} \label{subsec:sigma_v}
Both observations and models of the vertical structure of the atomic ISM suggest that the vertical pressure balance is maintained through turbulence \citep[e.g.,][]{LockmanGehman1991,KoyamaOstriker2009}. The varying velocity dispersions\footnote{In this work, we use ``velocity dispersion'' to refer to the cloud-to-cloud velocity dispersion, not the velocity dispersion within a single H\,{\scriptsize I} cloud.} of gas in discrete ISM phases contribute to the differences in vertical structure between phases, so quantifying the velocity dispersion of the CNM is important for constraining models of the multiphase structure of galaxies \citep[e.g.,][]{NarayanJog2002}. 

Table \ref{tab:sigma_v_fitting_results} lists the velocity dispersions, $\sigma_v$, inferred from the four fitting methods implemented by \texttt{kinematic\_scaleheight} for the same \bighicat{} subsamples discussed in Table \ref{tab:fitting_results}. The estimates for the least squares methods are systematically higher than those from the two MCMC methods. This is unsurprising given that the MCMC methods filter out outliers while the least squares method does not. We quote errors for the least squares methods using $\sigma_v/\sqrt{2N}$ to be consistent with \citet{Crovisier1978}. Errors for the MCMC methods are determined using a Bayesian approach \citep{Wenger2024}.

For the samples in Tables \ref{tab:fitting_results} and \ref{tab:sigma_v_fitting_results}, we derive $\sigma_v\sim5$--$8.5~\kms{}$.
\citet{McGDickey2007} investigated the kinematics of \hi{} observed in emission by the Southern Galactic Plane Survey. They found that \hi{} peculiar velocities were characterized by three components: a cold component with $\sigma_v=6.3~\kms{}$, a warmer component with $\sigma_v=12.3~\kms{}$, and a component with $\sigma_v=25.9~\kms{}$ (note, ``cold'' and ``warm'' here refer to the kinematics, not necessarily the actual temperature of the gas). Previously, \citet{LockmanGehman1991} derived a similar three-component fit for \hi{} emission at high latitudes. The velocity dispersions we report in Table \ref{tab:sigma_v_fitting_results} are consistent with the kinematically cold component identified in \hi{} emission observations. This consistency likely suggests that the component measured in emission is indeed tracing the CNM, which affirms that, at least in some cases, \hi{} emission can be used as a reliable tracer of the full multiphase atomic gas rather than just the warmer \hi{} \citep[e.g.,][]{Takakubo1967,Mebold1972,Haud2007,Marchal2019}.

As in Table \ref{tab:fitting_results}, we see differences in Table \ref{tab:sigma_v_fitting_results} between the results for optically thicker and higher column density samples --- the samples with $\tau>0.1$ or $N(\hi{})>10^{19.5}~\persc{}$ have velocity dispersions $\sim5.5~\kms{}$, whereas the samples with $\tau<0.1$ or $N(\hi{})<10^{19.5}~\persc{}$ have velocity dispersions $\sim8~\kms{}$. We note that the velocity dispersions measured for lower optical depth/lower column density subsamples are still in the range expected for CNM and lower than those expected for the WNM \citep[e.g.,][]{McGDickey2007,Murray2018}. The trend in $\sigma_v$ follows that observed in $\sigma_z$, with lower velocity dispersions inferred for samples of clouds with higher optical depths. This is consistent with the idea that the vertical balance of the cold gas is indeed maintained primarily by turbulence \citep{LockmanGehman1991,KoyamaOstriker2009}.

\begin{deluxetable*}{|c|c|c|c|c|} \label{tab:sigma_v_fitting_results}
\tablecaption{Estimates of $\sigma_v$ from the \texttt{kinematic\_scaleheight} code for different samples of \bighicat{} components. The first column describes each subsample. The second column lists the number of unique \bighicat{} components, $N$, belonging to each subsample. The third column lists the velocity dispersion derived using the \texttt{kinematic\_scaleheight} least-squares method, $\sigma_{v,\mathrm{ls}}$. The fourth column lists the velocity dispersion derived using the \texttt{kinematic\_scaleheight} moment ratio method, $\sigma_{v,\mathrm{MR}}$. The fifth column lists the velocity dispersion derived using the \texttt{kinematic\_scaleheight} shape method, $\sigma_{v,\mathrm{shape}}$.}
\tablehead{
\colhead{Sample} & \colhead{$N$} &  \colhead{$\sigma_{v,\mathrm{ls}}$} & \colhead{$\sigma_{v,\mathrm{MR}}$} & \colhead{$\sigma_{v,\mathrm{shape}}$} \\
\colhead{} & \colhead{} &  \colhead{\kms{}} & \colhead{\kms{}} & \colhead{\kms{}}
}
\startdata
    $|b|>5^{\circ}$ & 768 &  $11.7 \pm 0.3$ & $6.5 \pm 0.3$ & $6.5 \pm 0.4$ \\
    $\tau>0.1, |b|>5^{\circ}$ & 431  & $8.5 \pm 0.3$ & $5.8 \pm 0.4$ & $5.8 \pm 0.4$ \\
    $\tau<0.1, |b|>5^{\circ}$ & 336 & $14.3 \pm 0.6$ & $7.5 \pm 0.9$ & $7.5 \pm 0.9$ \\
    $N(\mathrm{HI})>10^{19.5}~\mathrm{cm}^{-2}, |b|>5^{\circ}$ & 429 & $9.0 \pm 0.3$ & $5.5 \pm 0.3$ & $5.5 \pm 0.3$ \\
    $N(\mathrm{HI})<10^{19.5}~\mathrm{cm}^{-2}, |b|>5^{\circ}$ & 216 & $14.1 \pm 0.7$ & $8.5 \pm 1.0$ & $8.5 \pm 1.0$ \\
    $T_{\mathrm{s}}<80~\mathrm{K}, |b|>5^{\circ}$ & 423 & $12.0 \pm 0.4$ & $6.4 \pm 0.4$ & $6.4 \pm 0.4$ \\
    $T_{\mathrm{s}}>80~\mathrm{K}, |b|>5^{\circ}$ & 257 & $10.2 \pm 0.4$ & $5.8 \pm 0.5$ & $5.8 \pm 0.5$ \\
    $\ts{}<80~\mathrm{K}, \tau>0.1, |b|>5^{\circ}$ & 296 & $8.8 \pm 0.4$ & $5.1 \pm 0.4$ & $5.1 \pm 0.4$ \\
    $\ts{}>80~\mathrm{K}, \tau<0.1, |b|>5^{\circ}$ & 171 & $11.2 \pm 0.6$ & $6.2 \pm 0.7$ & $6.2 \pm 0.7$ \\
\enddata
\end{deluxetable*}

\subsection{Constraints on the solar motion} \label{subsec:UVW}

Besides measuring the vertical thickness and velocity dispersion of the cold \hi{}, the methods employed in the \texttt{kinematic\_scaleheight} code also constrain the solar peculiar velocity components, which characterize the sun's motion relative to the LSR in the direction of the Galactic center ($U_{\odot}$), in the direction of rotation ($V_{\odot}$), and in the direction of the Galactic North pole ($W_{\odot}$). For the full sample of \bighicat{} features at $|b|>5^\circ$, we find $(U_{\odot},V_{\odot},W_{\odot})=(12.9\pm0.4,15.1\pm0.7,8.8\pm0.5)~\kms{}$. The results for the moment ratio and shape methods are nearly identical. Moreover, all of the samples in Table \ref{tab:fitting_results} have $U_{\odot}$, $V_{\odot}$, and $W_{\odot}$ values consistent with the results for the full sample. There is significant variation in the $U_{\odot}$, $V_{\odot}$, and $W_{\odot}$ values for different samples when estimated using the least squares approach. This is understandable given that the least squares approach is not robust to outliers, which make up $\sim10$--$30\%$ of each of the samples in Table \ref{tab:fitting_results}.

In recent years, a variety of techniques have been used to estimate the sun's peculiar motion (see Table 1 of \citealt{Ding2019} for a summary of results). Our estimates of $U_{\odot}$, $V_{\odot}$, and $W_{\odot}$ agree reasonably well (within $\lesssim3\sigma$) with some estimates derived from stellar kinematics \citep{Schonrich2010,Francis2013,FrancisAnderson2014,ZbindenSaha2019} and Galactic rotation models based on parallaxes \citep{Reid2014,Reid2019}. Nevertheless, estimates of $U_{\odot}$, $V_{\odot}$, and $W_{\odot}$ vary considerably depending on the observational sample and the method used, and our results are inconsistent with some measurements derived from stellar kinematics \citep[][and references therein]{Ding2019}.

\subsection{Inspection of outliers} \label{subsec:outliers}
\begin{figure*}
    \centering
    \includegraphics[width=\linewidth]{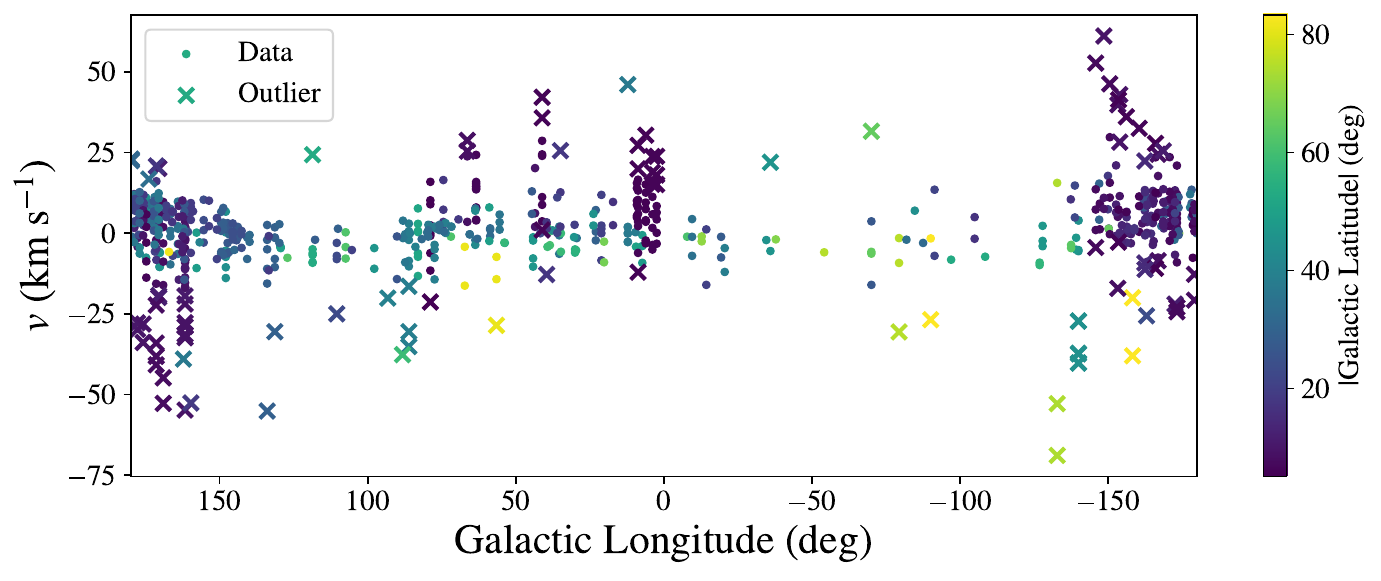}
    \caption{$\ell$-$v$ diagram of all 768 components in the \bighicat{} with $|b|>5^\circ$. Data are shown filled circles. Outliers are shown as X's. The color of each marker corresponds to the absolute value of the Galactic latitude of that component.}
    \label{fig:lv_w_outliers}
\end{figure*}

While we are primarily concerned with constraining the thickness of the cold atomic gas disk in the Milky Way, the identification of outliers in the Bayesian models developed by \citet{Wenger2024} allows us to investigate CNM structures that exist outside the standard kinematic and/or structural distribution. Outliers are identified as structures whose positions in $(\ell,b,v)$ space is inconsistent with our model of local gas clouds following circular Galactic rotation. Such structures could potentially trace dynamical events and inform our understanding of the formation and survival of the CNM in different environments. In most of the samples we consider, $\sim20\%$ of \hi{} structures are categorized as outliers. 
Figure \ref{fig:lv_w_outliers} shows the longitude-velocity ($\ell$-$v$) diagram for all the features in the $|b|>5^\circ$ sample. The outliers shown as X's and the non-outliers are shown as filled circles. Point are colored according to Galactic latitude.

\begin{figure}
    \centering
    \includegraphics[height=0.8\columnwidth]{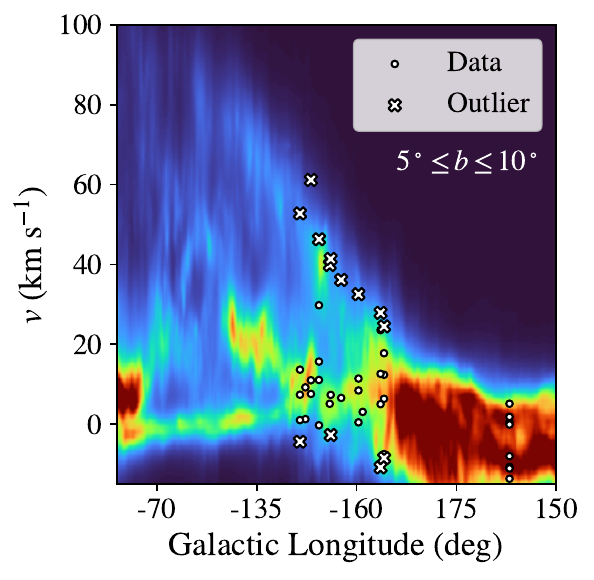}
    \includegraphics[height=0.8\columnwidth]{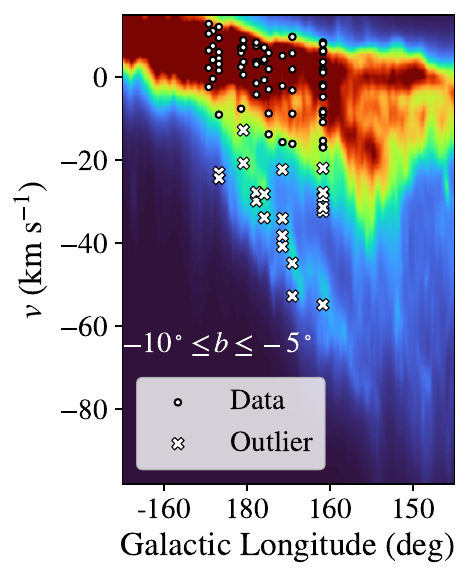}
    \caption{$\ell$-$v$ diagram showing components associated with flared/warped \hi{} in the outer Galaxy, associated with the Outer Arm \citep{Levine2006,KalberlaDedes2008}. White X's indicate points identified as outliers. White circles are the non-outlying components. The background map shows the \hi{} emission from \hi{}4PI \citep{HI4PI}, integrated over $5^\circ \leq b \leq 10^\circ$ (top) and $-10^\circ \leq b \leq -5^\circ$ (bottom).}
    \label{fig:OuterArm_lv}
\end{figure}

In Figure \ref{fig:OuterArm_lv}, we show that many of the outliers in the outer Galaxy are associated with the Outer Arm \citep[e.g.,][]{HouHan2014}. If we make a reasonable assumption about the disk thickness --- say, $\sigma_z\lesssim150~\mathrm{pc}$ \citep{Dickey2022,Wenger2024} --- then our latitude cut, $|b|>5^\circ$, ensures that a majority of the \hi{} structures in our sample are at a distance $<1~\mathrm{kpc}$ and virtually all of them are at a distance $<2~\mathrm{kpc}$. However, if the \hi{} disk in the outer Galaxy is warped \citep[][and references therein]{Binney1992} or flared \citep[][and references therein]{KalberlaKerp2009}, then our latitude criterion may not be sufficient to filter out features in the outer disk. Indeed, it is well-established that \hi{} towards the outer Galaxy (including the Outer Arm) is warped and flared \citep[e.g.,][]{Levine2006,KalberlaDedes2008}, so it is not surprising to have Outer Arm features interloping on our local \hi{} sample. Given that all of the features that are clearly associated with the Outer Arm are flagged as outliers, we still consider $|b|>5^\circ$ as a reasonable criterion for selecting local gas when using the Bayesian models from \citet{Wenger2024}.  On the other hand, for the least-squares method, which does not remove such outliers, a more stringent latitude cut is probably necessary to circumvent the features from the warped/flared outer Galaxy at intermediate latitudes.

\begin{figure*}[h!] 
\gridline{\fig{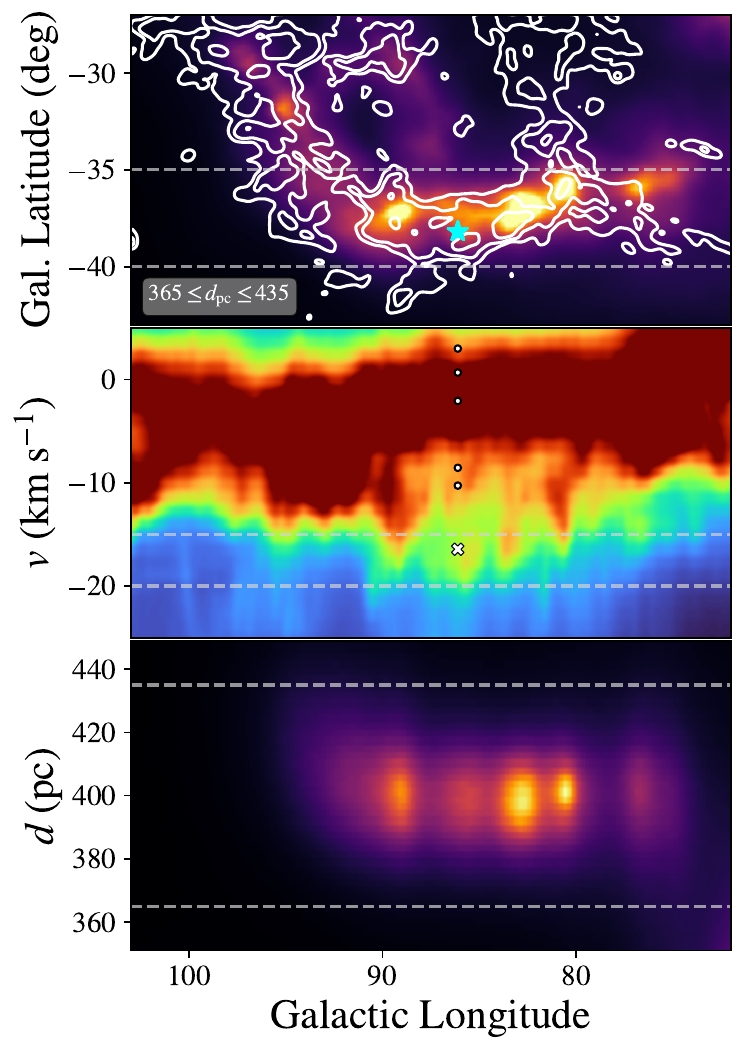}{0.34\textwidth}{3C454.3}
          \fig{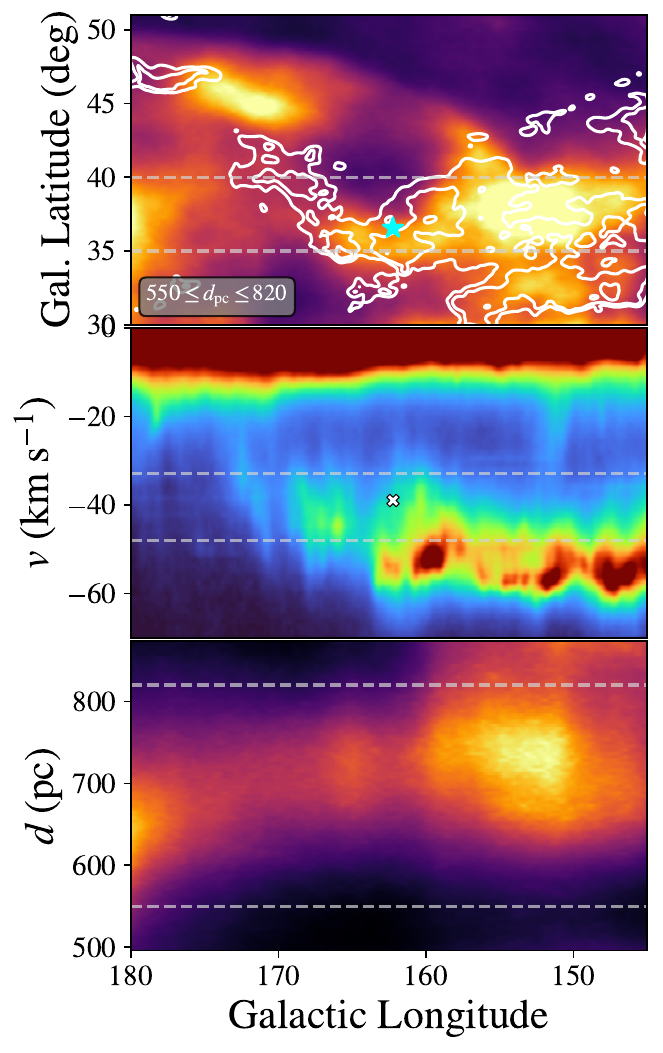}{0.3\textwidth}{J0834+555}
          \fig{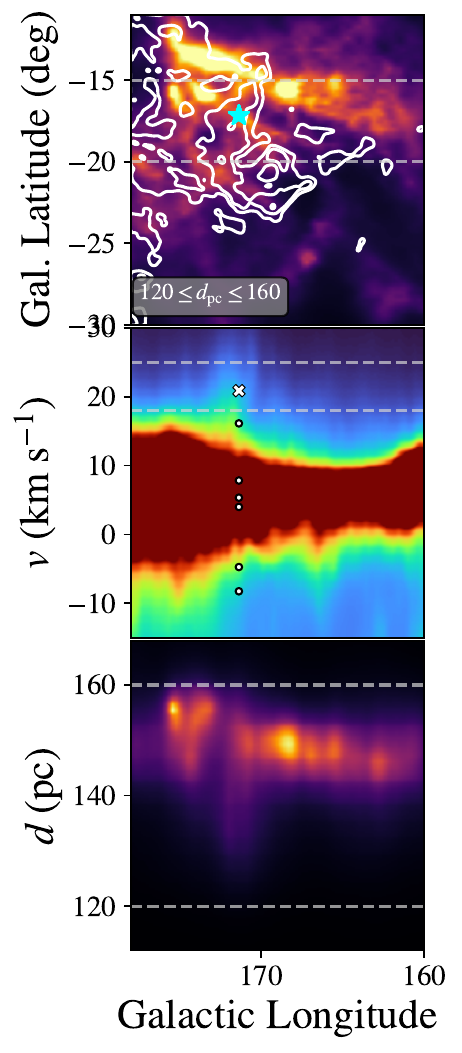}{0.21\textwidth}{4C+25.14}}
\gridline{\fig{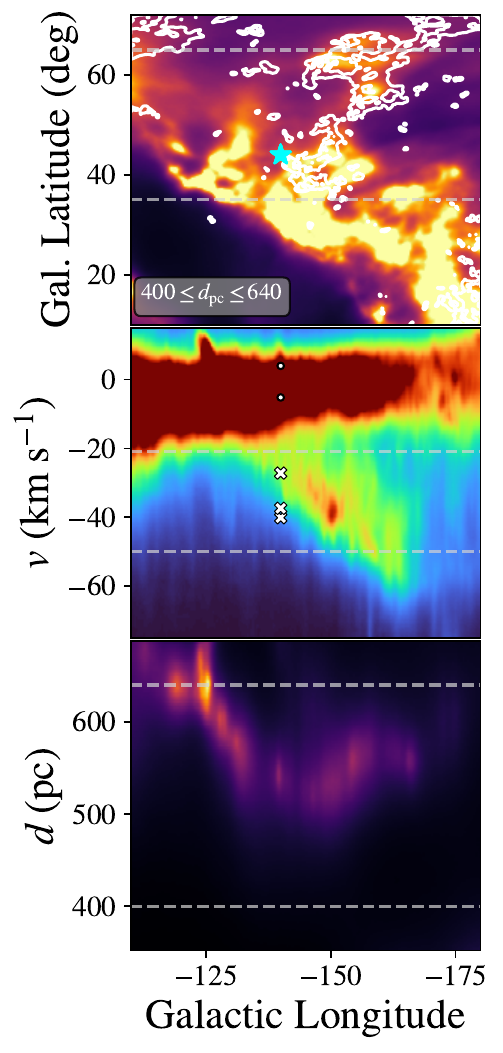}{0.235\textwidth}{3C225}
          \fig{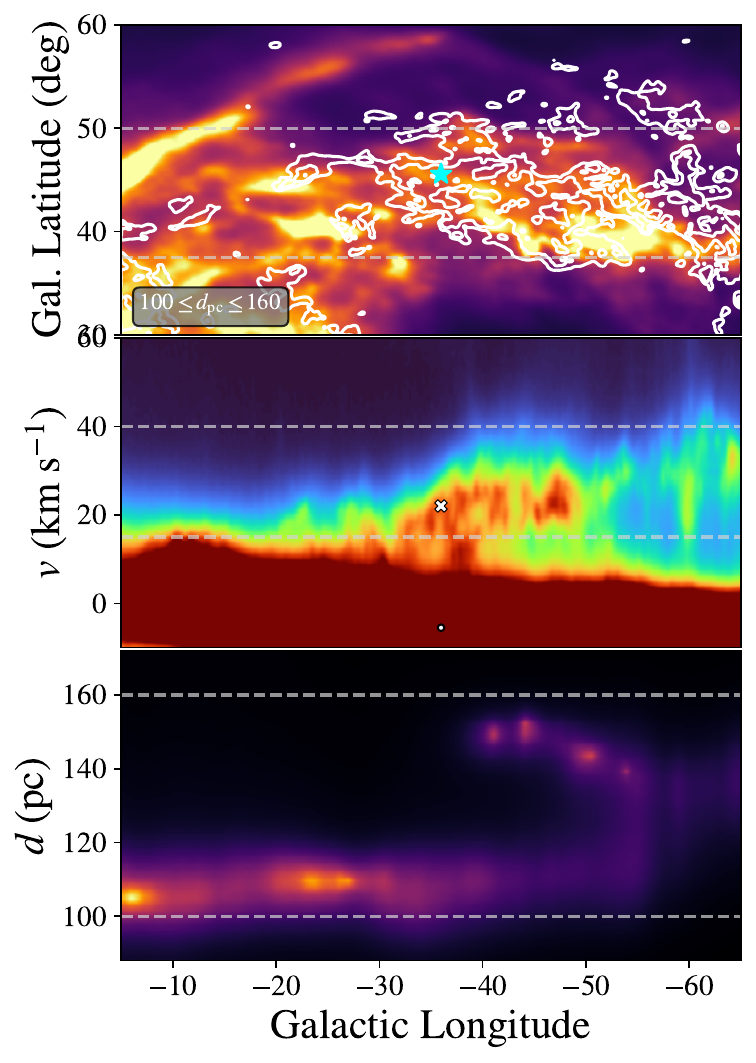}{0.35\textwidth}{J1351-148}
          \fig{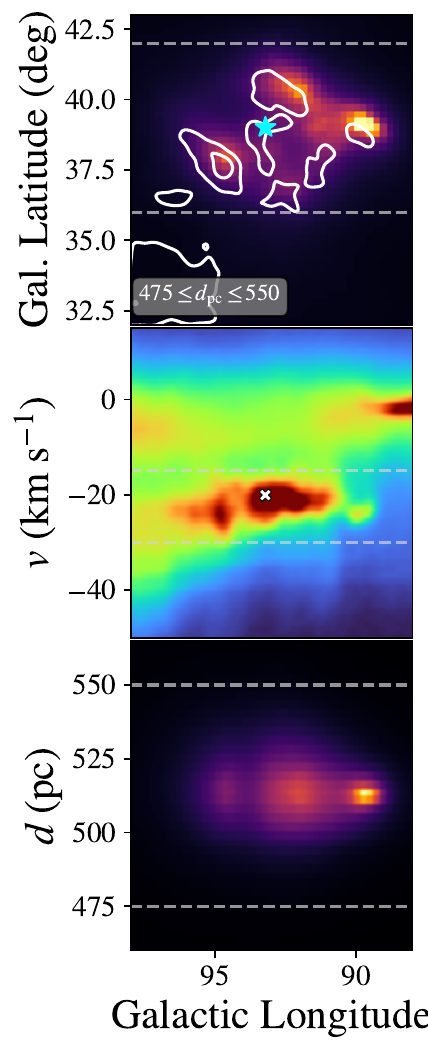}{0.2\textwidth}{J1638+625}}
\caption{Maps of gas associated with six outlying CNM components. \textit{Top panels:} \hi{} emission (white contours) overlaid on maps of dust emission. \hi{} emission is integrated over the velocity intervals shown with horizontal dashed line in the middle panels. Dust emission is integrated over the distance intervals shown with horizontal dashed lines in the bottom panels. \textit{Middle panels:} $\ell$-$v$ diagrams of \hi{} emission from \hi{}4PI \citep{HI4PI}. Outlying components are shown with white X's, non outlying components are shown as white circles. The \hi{} emission is integrated over the latitude intervals shown with horizontal lines in the top panels. \textit{Bottom panels:} $\ell$-$d$ diagrams from the \citet{Edenhofer2023} data cube. The dust emission is integrated over the latitude intervals shown with horizontal lines in the top panels.}
\label{fig:outlier_maps}
\end{figure*}

Still, some of outlying components in Figure \ref{fig:lv_w_outliers} indeed appear to trace local gas. 
Some outliers in the first quadrant have low latitudes, very likely tracing more distant gas structures (which are inconsistent with our model of the CNM at $d\lesssim~2~\mathrm{kpc}$). Meanwhile, other outliers have both high latitudes (indicating they are likely nearby) and high velocities. All of the velocities in the \bighicat{} sample are less than $62~\kms{}$, so the nearby outlying features do not represent high velocity clouds \citep[HVCs, $|v|>90~\kms{}$;][]{WakkerVanWoerden1997}, but they may represent intermediate velocity clouds (IVCs).
To characterize these gas structures --- and perhaps better understand the origin of IVCs --- we try to identify counterparts in \hi{} emission and 3D space.
In Figure \ref{fig:outlier_maps}, we present maps of \hi{} emission (top panel, contours) and dust emission (top panel, background) associated with six components that are identified as outliers.
These six components were selected based on two criteria: (1) we were able to identify \hi{} emission that was spatially and kinematically associated with the \hi{} absorption; and (2) we were further able to connect the \hi{} emission structures to structures in the recent three-dimensional map of the ISM constructed by \citet{Edenhofer2023}.

To illustrate the association between these six \hi{} absorption components with the \hi{} emission, we also show $\ell$-$v$ diagrams of the \hi{} emission in these directions (middle panels in Figure \ref{fig:outlier_maps}).
As before, the \bighicat{} features flagged as outliers are overlaid as X's and features not flagged as outliers are overlaid as circles We isolate the \hi{} emission at velocities associated with the outlying components (outlined by horizontal dashed lines in the $\ell$-$v$ diagrams). The emission at these velocities is shown in the white contours in the top panels of Figure \ref{fig:outlier_maps}.
We then identified structures in the \citet{Edenhofer2023} dust maps that showed a similar morphology to the \hi{} emission (the background images in the top panels of Figure \ref{fig:outlier_maps}, identified by eye).
In the bottom panels of Figure \ref{fig:outlier_maps}, we also show longitude-distance ($\ell$-$d$) diagrams (similar to $\ell$-$v$ diagram, but the $y$-axis is a third spatial dimension, rather than a spectral one) in these directions. We use horizontal dashed lines to show the distances to the dust structures that appear to be associated with the structures identified in \hi{} emission. In this way, we characterized the kinematics and morphology of \hi{} emission and 3D density structures of a selection of outlying components.

Five of the six components (illustrated in the first five panels in Figure \ref{fig:outlier_maps}, in the directions of 3C454.3, J0834+555, 4C+25.14, J1638+625, and J1351-148) appear to be associated with extended arc-like  features. The structure in the direction of 4C+25.14 appears to be associated with the local bubble wall \citep{ONeill2024}.  
The \hi{} emission in the direction of J1351-148 at $\sim20$--$40~\kms{}$ is associated with the local bubble wall (clear in the $\ell$-$d$ diagram; \citealt{ONeill2024} report a distance to the local bubble wall of $107~\mathrm{pc}$ in this direction) as well as an extended structure at a distance of $\sim110$--$160~\mathrm{pc}$ on the eastern side (see the $\ell$-$d$ diagram at $-60^\circ \lesssim\ell\lesssim -40^\circ$). While both of these components contribute to the \hi{} emission at these velocities, at the longitude of J1351-148, it appears that only the nearer structure (i.e., local bubble wall) is present, so we conclude that this outlying component is associated with the local bubble wall.
The other four structures shown in Figure \ref{fig:lv_w_outliers} are not associated with the local bubble wall, but are instead associated with discrete structures $400$--$600~\mathrm{pc}$ from the sun.

The analysis here remains limited, but future work connecting cold \hi{} structures at anomalous velocities (perhaps representing IVCs) with the 3D dustmaps in a more systematic way \citep[see, e.g.,][]{Soler2023} could help establish the origin of these cold gas structures. For example, here we find that two CNM clouds with anomalous velocities are associated with the local bubble wall \citep{Welsh2004}. More broadly, such comparisons could also help illuminate the structure of the CNM and its relation to the disk-halo interface \citep{Lockman2002,SS2006}.

\section{Discussion} \label{sec:discussion}

Consistent with \citet{Wenger2024}, we find that the vertical thickness of the CNM in the solar neighborhood, $\sigma_z$, is significantly smaller than previously reported \citep{Crovisier1978,BelfortCrovisier1984,Dickey2022}. We further find that the derived thickness depends strongly on the physical properties of the \hi{} structures considered --- \hi{} structures with optical depths $\gtrsim0.1$ and/or column densities $\gtrsim10^{20}~\persc{}$ are characterized by a scale height $\sigma_z\approx50$--$60~\mathrm{pc}$, while \hi{} structures with optical depths $\lesssim0.1$ and/or column densities $\lesssim10^{20}~\persc{}$ are characterized by a scale height $\sigma_z\approx100$--$120~\mathrm{pc}$. It is worth noting here that the sensitivity of the \citet{Crovier1978_survey} \hi{} absorption survey means that \citet{Crovisier1978} was considering primarily structures with $\tau\gtrsim0.1$. Thanks to improvements in sensitivity over the past half century, approximately half of the structures in the \bighicat{} would not have been detectable by \cite{Crovier1978_survey} and so would not have been included in the sample of \citet{Crovisier1978}.

The results presented here challenge the idea that the CNM thickness increases monotonically (with local variations) with Galactocentric radius \citep[e.g.,][]{KalberlaDedes2008,Dickey2022,McG2023}. Instead, they imply that the thickness of the CNM is roughly the same at the solar circle as in the inner Galaxy, at least for $\tau\gtrsim0.1$. 

Recently, \citet{Smith2023} investigated the CNM distribution at high-resolution (physical scales $\lesssim1~\mathrm{pc}$) in a simulation of an isolated spiral galaxy \citep[updated from][]{Tress2020,Tress2021}. They showed that the thickness of the CNM was systematically narrower than that of the total \hi{} (CNM+WNM) and was roughly constant across most of the disk (excluding the center, where $\sigma_z$ was smaller), with $\sigma_z\sim100~\mathrm{pc}$. While the value of $\sigma_z$ measured for the CNM was mostly stable, significant ($\sim$factor of two) local variations did appear in some parts of the galaxy. They considered both a constant radiation field and a radially decreasing radiation field; the radial extent of the CNM changed between the two models, but the vertical distribution of the CNM was similar (and similarly flat with Galactocentric radius) in both models. 
Their results are consistent with the picture developed in this work, where the vertical thickness of the CNM in the solar neighborhood is similar to that in the inner Galaxy \citep{Dickey2022} (compare Figure \ref{fig:araa_comparison} to Figure 10 of \citealt{Smith2023}).

The vertical equilibrium of the CNM  
is set by the balance of gravity and the pressure gradient. In the CNM, it is thought that the turbulent pressure alone may be sufficient to sustain the vertical equilibrium \citep{LockmanGehman1991,KoyamaOstriker2009}; the thermal pressure is minimal for such cold gas but more important for warmer ISM components \citep[e.g.,][]{KimOstriker2015}. Indeed, in both observations and simulations of galaxies, the pressure of the cold atomic ISM is dominated by dynamical processes \citep{KimOstriker2015,HerreraCamus2017}. 
While our results imply a thinner CNM disk than reported by some previous experiments \citep{Crovisier1978,BelfortCrovisier1984,LockmanGehman1991,Dickey2022}, Table \ref{tab:sigma_v_fitting_results} also shows that the Bayesian models from \citet{Wenger2024} infer relatively small velocity dispersions (see the two rightmost columns). For example, the velocity dispersions in Table \ref{tab:sigma_v_fitting_results} are about a factor of two smaller than those measured by \citet{LockmanGehman1991} from \hi{} emission. The scaleheights in Table \ref{tab:fitting_results} are also about a factor of two smaller than those inferred by \citet{LockmanGehman1991}, so despite our updated estimates of $\sigma_z$ and $\sigma_v$, it remains plausible that the vertical support of the CNM can still be explained by turbulence.

The CNM precedes the formation of molecular clouds \citep[e.g.,][]{Goldsmith2007,Stanimirovic2014}, so the structure of the CNM influences the structure of the molecular gas disk.
The molecular gas traced by CO in the Milky Way is known to be organized in a thin disk, $\sigma_z\sim40~\mathrm{pc}$, and a thick disk, $\sigma_z\sim120~\mathrm{pc}$ \citep{DameThaddeus1994,Malhotra1994,Su2021}. In Figure \ref{fig:araa_comparison}, estimates for the total molecular gas disk are shown as opaque yellow squares \citep[see Figure 6 of ][and references therein]{HeyerDame2015}, while estimates for the thin and thick components are shown as semi-transparent yellow squares \citep{Su2021}.
The thickness of the thin disk is similar to our estimate for the \hi{} structures with $\tau>0.1$ (especially if we only consider those with $\ts<80~\mathrm{K}$; see Table \ref{tab:fitting_results}). 
The similarity of the vertical thickness of the molecular gas disk with the vertical thickness we find for cold, optically thick \hi{} is consistent with recent work that suggests that only this \hi{} contributes significantly to the formation of molecular clouds \citep{Stanimirovic2014,Nguyen2019,Ryb22,Park2023,Hafner2023}. The connection is particularly striking because we find that the \hi{} in this thin disk has $\tau\gtrsim0.1$ and $N(\hi{})\gtrsim10^{20}~\persc{}$, which are roughly the same criteria that have been deemed necessary for the formation of molecular gas \citep{Ryb22,Park2023,Hafner2023}. It is also worth noting that the thickness of the molecular gas disk has been shown to be roughly constant with Galactocentric radius from 
$R\sim 3~\mathrm{kpc}$ out to the solar circle \citep[][and references therein]{HeyerDame2015}, consistent with the trend that Figure \ref{fig:araa_comparison} suggests exists for cold, optically thick \hi{}.

Meanwhile, the thickness of the thick molecular disk, $\sigma_z\sim120~\mathrm{pc}$, is similar to our estimate for the thickness of the atomic gas disk for absorbing \hi{} structures with $\tau<0.1$ (see Figure \ref{fig:araa_comparison}).
Recently, \citet{Ryb22} identified diffuse molecular gas (traced by HCO$^+$ absorption) associated with the CNM at optical depths $<0.1$. The spectral signature of this diffuse gas bore striking resemblance to a signature identified in OH emission by \citet{Busch2021}, who showed that this diffuse gas belonged to a thick disk (they used a very simple cylindrical model with a thickness of $200~\mathrm{pc}$). The similarity between our measurement of the optically thinner CNM and the thick molecular disk \citep{DameThaddeus1994,Malhotra1994,Su2021}, as well as the recent detection of diffuse molecular gas associated with this \hi{}, could indicate that some of the molecular gas and the cold atomic gas at higher $z$ share a common origin.

\section{Conclusions} \label{sec:conclusions}

We present new estimates of the vertical thickness of the cold atomic gas disk in the solar neighborhood. We apply the \texttt{kinematic\_scaleheight} code \citep{Wenger2024,Wenger2024_code}, which uses four methods --- including the method developed by \citet{Crovisier1978} --- to estimate the vertical thickness of the disk from a given sample of absorbing structures, to subsamples of the \bighicat{}. We focus only on \hi{} structures at Galactic latitudes $|b|>5^\circ$, where most of the gas along the line of sight is local ($d<2~\mathrm{kpc}$).

As discussed in \citet{Wenger2024}, the three new approaches to measure the vertical thickness of the CNM disk in the solar neighborhood have corrected the \citet{Crovisier1978} approach and lead to a smaller disk height than previously reported \citep{Crovisier1978,BelfortCrovisier1984,Dickey2022}. In fact, the values of $\sigma_z$ that we measure for different samples of local \hi{} structures (Table \ref{tab:fitting_results}) are similar to those measured in the inner Galaxy, at $R\sim3~\mathrm{kpc}$ \citep{Dickey2022}. This challenges the conventional picture of a vertical thickness that gradually rises with Galactocentric radius \citep[][see Figure \ref{fig:araa_comparison}]{McG2023}. Instead, this is more in line with recent simulations of the multiphase ISM \citep{Smith2023} that have found the CNM thickness is roughly constant with Galactocentric radius, from just outside the Galactic center out to the outer galaxy \citep[although they do not find CNM at very large galatctic radii, as measured in observations of the Milky Way from ][]{Dickey2009}.

If we consider the entire \bighicat{} sample at $|b|>5^\circ$, we find $\sigma_z\approx75\pm8~\mathrm{pc}$ (assuming a Gaussian shape to the vertical distribution).
However, if we divide the high-latitude \bighicat{} into an optically thicker sample ($\tau>0.1$) and an optically thinner sample ($\tau<0.1$), we find that the optically thicker sample has a significantly narrower vertical distribution than the optically thicker sample, with $\sigma_z\approx60~\mathrm{pc}$ and $\sigma_z\approx120~\mathrm{pc}$, respectively. We find similar results if we separate by column density (with a boundary of $N(\hi{})\sim10^{20}$) instead of optical depth. This is noteworthy because $\tau\approx0.1$ and $N(\mathrm{HI})\approx10^{19.5}$ have been shown to be necessary (but not sufficient) prerequisites for molecule formation in the diffuse ISM. The thicknesses we derive for the optically thicker or higher column density samples, $\sigma_z\sim50$--$60~\mathrm{pc}$, are consistent with estimates for the thickness of the molecular gas disk in the solar neighborhood \citep[e.g.,][]{Sanders1984,Grabelsky1987,Bronfman1988,Clemens1988,Malhotra1994,Nakanishi2006}. Our results are consistent with the interpretation that optically thick, high column density CNM gas leads to the formation of molecular clouds \citep[e.g.,][]{Stanimirovic2014}.

Nevertheless, we do not find significant variations in the disk thickness as a function of \hi{} spin temperature (approximately equal to the gas kinetic temperature in these environments).
However, \citet{Hill2018} showed that, in 3D magnetohydrodynamic simulations of the multiphase ISM \citep{JoungMacLow2006,Joung2009,Hill2012}, the distribution of gas at temperatures $\lesssim200~\mathrm{K}$ was relatively uniform for $|z|\lesssim100~\mathrm{pc}$. 89\% of the components in our high-latitude \bighicat{} sample have $\ts<200~\mathrm{K}$. Variations in the scaleheight as a function of temperature are thought to manifest over much larger temperature ranges than considered here \citep[see Figure 6 of][]{Hill2018}.

For each sample of \hi{} structures we inspect, we estimate both the vertical thickness of the cold disk, $\sigma_z$, as well as the cloud-to-cloud velocity dispersion, $\sigma_v$. The velocity dispersions are reported in Table \ref{tab:sigma_v_fitting_results}, ranging from $5.1~\kms{}$ to $8.5~\kms{}$. These velocity dispersions are significantly lower than those inferred using the original technique implemented by \citet{Crovisier1978} (see the third column of Table \ref{tab:sigma_v_fitting_results}) as well as that which \citet{LockmanGehman1991} inferred from \hi{} emission to estimate the thickness of the cold disk. Yet, as shown in Table \ref{tab:fitting_results}, our estimates of the disk thickness are also lower. If we make a reasonable estimate of the gravitational potential close to the plane (e.g., \citealt{KuijkenGilmore1989}, as discussed by \citealt{LockmanGehman1991}), then we find that the inferred velocity dispersions (Table \ref{tab:sigma_v_fitting_results}) and the inferred scaleheights (Table \ref{tab:fitting_results}) remain consistent with the vertical balance of the CNM being maintained primarily by turbulence \citep{LockmanGehman1991,KoyamaOstriker2009}.

Besides measuring the vertical distribution and velocity dispersion of cold \hi{}, we are also able to constrain the solar motion with respect to the LSR. For the full high-latitude \bighicat{} sample, we find $(U_{\odot},V_{\odot},W_{\odot})=(12.9\pm0.4,15.1\pm0.7,8.8\pm0.5)~\kms{}$, consistent with estimates derived from stellar kinematics \citep{Schonrich2010,Francis2013,FrancisAnderson2014,ZbindenSaha2019} and Galactic rotation models based on parallaxes \citep{Reid2014,Reid2019} \citep[although estimates of $U_{\odot}$, $V_{\odot}$, and $W_{\odot}$ depend on both the observational sample and the method used, and our results are inconsistent with some other measurements; ][and references therein]{Ding2019}.

We also identify outliers that are not well-fit by the simple Galactic rotation model; some of these outliers may represent IVCs. In Section \ref{subsec:outliers}, we show that several of these outlying components are associated with extended arc-like structures. 
Further, the recent emergence of 3D maps of the local ISM \citep[e.g.,][]{Lallement2019,Green2019,Leike2020,Edenhofer2023} has made it possible to map the 3D spatial position and morphology of gas structures identified in two spatial dimensions using spectral data cubes \citep[see, e.g.,][]{Soler2023}. Here, we find that two of the cold \hi{} outliers are very likely associated with the local bubble wall \citep{ONeill2024}.

\newpage
\noindent
This research has made use of NASA’s Astrophysics Data System Bibliographic Services.
D.R.R. is supported by a National Science Foundation Astronomy and Astrophysics Postdoctoral Fellowship under award AST-2303902.
T.V.W. is supported by a National Science Foundation Astronomy and Astrophysics Postdoctoral Fellowship under award AST-2202340.  
The authors acknowledge Interstellar Institute's program ``II6'' and the Paris-Saclay University's Institut Pascal for hosting discussions that nourished the development of the ideas behind this work.
We would also like to thank N. M. McClure-Griffiths, R. Benjamin, and J. M. Dickey for useful discussions regarding the content of this paper.

\software{Astropy \citep{Astropy2013}, \texttt{kinematic\_scalehieght} \citep{Wenger2024_code}.}

\bibliography{refs}{}
\bibliographystyle{aasjournal}

\end{document}